\documentclass[11pt]{article}

\usepackage[preprint]{acl}
\usepackage{times}
\usepackage{latexsym}
\usepackage[T1]{fontenc}
\usepackage[utf8]{inputenc}
\usepackage{microtype}
\usepackage{inconsolata}
\usepackage{graphicx}
\usepackage{booktabs}
\usepackage{amsmath}
\usepackage{amssymb}
\usepackage{multirow}
\usepackage{xcolor}
\usepackage{algorithm}
\usepackage{algorithmic}

\newtheorem{proposition}{Proposition}

\newcommand{\sunrel}{s_k^{\text{unrel}}}
\newcommand{\sclone}{s_k^{\text{clone}}}

\title{Do System Prompts Leave Behavioral Fingerprints? \\
A Large-Scale Empirical Study of Clone Detection via Output Similarity}

\author{%
  \textbf{Linghan Chen}\textsuperscript{1*} \quad
  \textbf{Yudong Gao}\textsuperscript{2*} \quad
  \textbf{Jiyao Wang}\textsuperscript{1} \quad
  \textbf{Kaiyan Ji}\textsuperscript{1} \quad
  \textbf{Honglong Chen}\textsuperscript{3\textdagger} \\[4pt]
  \textsuperscript{1}University of Adelaide \quad
  \textsuperscript{2}Hong Kong University of Science and Technology \\
  \textsuperscript{3}China University of Petroleum \\[4pt]
  \texttt{linghanchen2004@gmail.com}
}

\begin{document}
\maketitle

{\renewcommand{\thefootnote}{*}\footnotetext{Equal contribution.}}
{\renewcommand{\thefootnote}{\textdagger}\footnotetext{Corresponding author.}}

\begin{abstract}
System prompts can be extracted from commercial LLMs with over 80\% success and redeployed at zero cost, yet a prompt owner has no way to verify whether a suspected deployment is a clone. We propose Black-Box Behavioral Fingerprinting (BBF): the prompt owner registers a behavioral signature from model outputs and later tests whether a suspect deployment matches that signature more closely than an unrelated baseline. BBF requires only black-box API access. Through a large-scale study (4 model families, 8 benchmarks, 288,000 responses), we find that prompt choice explains 24.4\% of output variance and same-model detection reaches AUC 0.876. Cross-model performance is bounded by detector identity, with off-diagonal AUC ranging from 0.845 (Claude as detector) down to 0.665 (Qwen) and overall mean 0.725. BBF resists non-adaptive prompt paraphrasing (AUC $\geq 0.889$) and is robust to imperfect extraction, but a single-sentence formal-tone prefix can collapse detection on short structured outputs (MNLI 0.978 $\to$ 0.547), isolating style-invariant detection as the key open problem. Diagnostic Query Optimization, a zero-cost query selection rule, adds +0.120 to cross-model AUC.
\end{abstract}

\section{Introduction}
\label{sec:intro}

LLM application platforms such as OpenAI's GPT Store, Coze, and Dify have given rise to a new class of intellectual property: the system prompt. A carefully crafted prompt can transform a general-purpose model into a specialized product by encoding domain expertise, reasoning strategies, and output conventions, and LLM-based agents extend this further by encoding tool-use workflows, multi-step reasoning chains, and safety guardrails \citep{wang2024survey}. Our study covers the single-turn case directly; the multi-turn agent setting is a natural extension we discuss in Limitations. Such prompt-based applications are now widely deployed commercially, and the prompt is often a key differentiator between competing products. However, unlike model weights, which require significant compute to replicate, a system prompt can be extracted verbatim through adversarial queries, with success rates exceeding 80\% on commercial systems \citep{hui2024pleak, sha2024prompt}. Once extracted, a prompt can be redeployed on any model, on any platform, at zero cost. \citet{yao2025prsa} showed that this entire pipeline, from extraction to cloning to deployment, can be fully automated against real-world prompt services.

Existing defenses focus on \emph{preventing} extraction \citep{schulhoff2023ignore, chen2025proxyprompt, das2025system}, yet the extraction success rates above show that prevention alone is not yet reliable. A single successful extraction also permanently compromises the prompt, since the extracted text can be copied and redistributed without limit. We therefore argue for a complementary approach: \emph{detecting} whether a suspected deployment is running a stolen prompt. Just as copyright enforcement provides recourse even when access control fails, post-extraction detection offers recourse even after the prompt has already been stolen.

\begin{figure}[t]
\centering
\includegraphics[width=\columnwidth]{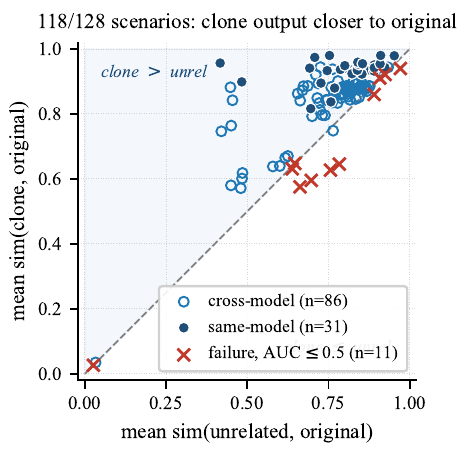}
\caption{Each point is one of 128 scenarios (8 benchmarks $\times$ 4 registration $\times$ 4 detection models). Red $\times$ marks failures (AUC $\leq 0.5$).}
\label{fig:scatter}
\end{figure}

Whether such detection is feasible hinges on a prior question: does a system prompt leave a usable trace in model outputs at all? We find that it does, and strongly: prompt choice explains 24.4\% of output-embedding variance across eight benchmarks ($\eta^2 = 0.244$, a large effect \citep{cohen1988}; Section~\ref{sec:eta}), and clone outputs sit closer to the original than unrelated-prompt outputs do in 118 of 128 experimental scenarios (Figure~\ref{fig:scatter}). This imprint is what makes post-extraction detection feasible in principle; the open questions concern its \emph{robustness}. Does it survive \emph{cross-model transfer} to a different LLM with different tokenization and decoding? Does it survive an adversary who \emph{paraphrases} the stolen prompt? Which \emph{domains and model pairs} are boundary conditions, and how many \emph{diagnostic queries} does a defender need?

We propose Black-Box Behavioral Fingerprinting (BBF), inspired by content fingerprinting in multimedia: the prompt owner registers a behavioral signature from outputs of the original deployment, and later tests whether a suspect produces outputs closer to that signature than to an unrelated baseline. BBF needs only black-box API access.

Through a large-scale empirical study (128 scenarios, 4 model families, 8 benchmarks, 288,000 responses), we make three contributions. First, we map when output-similarity detection succeeds and when it fails across the full $4 \times 4 \times 8$ space: same-model AUC = 0.876, cross-model AUC ranges 0.845 (Claude as detector) to 0.665 (Qwen as detector), and 11 of 128 scenarios fail along three identifiable patterns (short outputs, data artifacts, model-pair style mismatch; Section~\ref{sec:results}). Second, we propose Diagnostic Query Optimization (DQO), a prompt-free and transferable query selection rule with mean held-out gains of +0.120 cross-model and +0.045 same-model (Section~\ref{sec:dqo}). Third, across Sections \ref{sec:ablation}--\ref{sec:robustness}, we show that BBF survives heavy prompt paraphrasing (AUC $\geq 0.889$), characterize one adaptive attack that does break it on short-output tasks, and establish that 25 diagnostic queries already suffice for reliable detection.

BBF is a \emph{verification} tool, not a discovery tool: it confirms or refutes suspicion about a specific deployment and is intended as a complement to existing IP enforcement, not a surveillance method.

\section{Related Work}
\label{sec:related}

\paragraph{System prompt extraction.} System prompts can be extracted through adversarial queries \citep{schulhoff2023ignore, hui2024pleak, sha2024prompt}, indirect prompt injection \citep{greshake2023not, toyer2024tensor}, and automated attacks against real-world platforms \citep{yao2025prsa}. Defenses include proxy architectures \citep{chen2025proxyprompt} and various filtering strategies surveyed by \citet{das2025system}. All of these focus on \emph{preventing} extraction, whereas our work targets \emph{post-extraction detection} as a complementary defense layer.

\paragraph{Watermarking and model fingerprinting.} LLM watermarking \citep{kirchenbauer2023watermark, christ2024undetectable} embeds a signal into model outputs during generation, requiring modification of the decoding process and model provider cooperation. Model fingerprinting \citep{xu2024instructional, zeng2024fingerprint, sun2025rofl, li2023modeldiff} identifies which \emph{model} produced an output, typically requiring weight access or fine-tuning; prompts have also been used as units of personalization in adjacent privacy contexts such as federated unlearning \citep{wu2025mimir}. Our work differs along two axes: we detect whether a specific \emph{system prompt} has been cloned (not which model is used), and we require only black-box API access with no weights, no generation modification, and no platform cooperation. Direct numerical comparison against these methods is not meaningful given the different access assumptions; the closest output-only baseline is lexical similarity (TF-IDF / Jaccard), which we benchmark in Section~\ref{sec:embedding_ablation}.

\paragraph{Sentence embeddings.} BBF builds on sentence-level representations \citep{reimers2019sentencebert, li2023towards, wang2024e5} as benchmarked by MTEB \citep{muennighoff2023mteb}. These embeddings serve as a measurement tool; our contribution is establishing when the resulting similarity signal suffices for clone detection. A natural alternative would be to ask a strong LLM to judge whether two output sets share a prompt (LLM-as-judge); we did not adopt this baseline because each detection call would invoke a frontier model on the full diagnostic set (cost scales with $K \cdot N$), the judgments are themselves prompt-conditioned and non-deterministic, and embedding similarity gives a closed-form, reproducible decision rule. A direct comparison against an LLM-as-judge baseline is a useful next experiment.

\begin{figure*}[t]
\centering
\includegraphics[width=\textwidth]{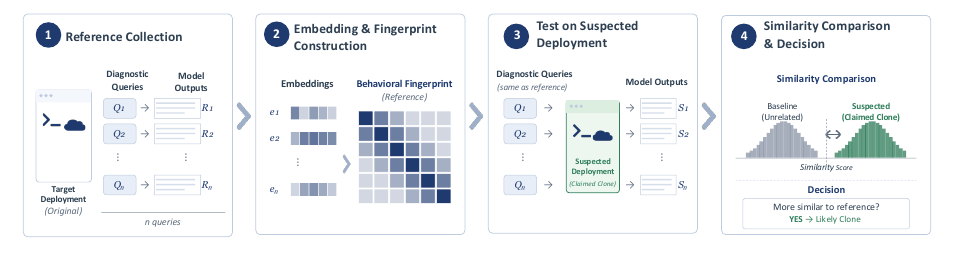}
\caption{BBF at a glance. \textbf{(1)} The defender sends $n$ diagnostic queries to the original deployment and collects responses. \textbf{(2)} Responses are embedded and aggregated into a behavioral fingerprint that characterizes the original prompt's output distribution. \textbf{(3)} The same queries are sent to the suspected deployment. \textbf{(4)} The suspect's outputs are compared in embedding space to both the reference fingerprint and an unrelated-prompt baseline; greater similarity to the reference indicates cloning. Formal definitions in Section~\ref{sec:method}.}
\label{fig:pipeline}
\end{figure*}

\section{Method}
\label{sec:method}

\subsection{Problem Formulation}
\label{sec:formulation}

Let $\mathcal{P}$ be a protected system prompt and $\mathcal{M}$ an LLM. For query $q$, the model produces a stochastic output $o \sim \mathcal{M}(q; \mathcal{P})$. A sentence encoder $\phi : \mathcal{O} \to \mathbb{R}^d$ maps outputs into a normalized embedding space. The prompt $\mathcal{P}$ thus induces a query-conditional distribution $p_{\mathcal{P}}(\cdot \mid q)$ over $\mathbb{R}^d$. An adversary extracts $\mathcal{P}$, creates a clone $\mathcal{P}'$, and deploys it on a potentially different model $\mathcal{M}'$. The clone detection problem is: given only black-box API access to both deployments, determine whether the behavioral distribution $p_{\mathcal{P}'}(\cdot \mid q)$ is closer to $p_{\mathcal{P}}(\cdot \mid q)$ than to that of an unrelated prompt $\mathcal{P}_u$.

Two prompts are \emph{functionally equivalent} if they elicit comparable task behavior; the clones we evaluate are semantic paraphrases of their originals that we manually verified to preserve task instructions on spot-checked benchmarks. BBF operationalizes detection of this equivalence as a centroid-closeness test in embedding space, comparing the suspect's centroid against the original and an unrelated same-domain baseline.

\subsection{Threat Model}
\label{sec:threat}

The detection setting involves two parties with asymmetric capabilities. The \emph{defender} (prompt owner) can query the suspected clone's public API, send $\geq$25 diagnostic queries, and compute a baseline using one unrelated prompt from the same domain. The defender does not need to know which model the suspect uses, does not need model weights or logits, and does not need cooperation from the suspect's platform. One assumption is implicit: the defender must know the protected prompt's broad domain in order to construct a same-domain unrelated baseline. This is mild in practice (a prompt owner who suspects cloning typically knows what their own prompt does), but it is a real prior, and misspecifying the domain weakens the baseline. Section~\ref{sec:deployment} reports a leave-one-domain-out cross-validation in which the baseline is constructed from a different benchmark, yielding mean AUC 0.793 versus 0.876 in-domain, which quantifies the cost of imperfect domain knowledge. The \emph{adversary} may deploy the stolen prompt on any model family, may paraphrase the prompt to obscure its origin (evaluated in Section~\ref{sec:robustness}), and may use prompts obtained through imperfect extraction rather than verbatim copies (evaluated in Section~\ref{sec:deployment}). This asymmetry, where the defender has minimal access while the adversary has maximal freedom, makes the detection problem challenging and motivates a method that relies on behavioral signals rather than direct comparison of the prompt text.

\subsection{BBF: Black-Box Behavioral Fingerprinting}
\label{sec:bbf}

BBF is intentionally simple: it casts clone detection as a two-sample comparison in embedding space, using only off-the-shelf sentence embeddings and cosine similarity. This design choice isolates the core empirical question (do behavioral fingerprints exist and transfer?) from confounds that a more complex method would introduce. BBF operationalizes detection through two phases.

\paragraph{Registration.} For diagnostic queries $\mathcal{Q} = \{q_1, \ldots, q_K\}$, the defender draws $N$ i.i.d.\ output samples from the original deployment and estimates the behavioral centroid $\boldsymbol{\mu}_k$ for each query:
\begin{equation}
\hat{\boldsymbol{\mu}}_k = \frac{1}{N} \sum_{i=1}^{N} \phi\big(\mathcal{M}(q_k; \mathcal{P})_i\big).
\label{eq:centroid}
\end{equation}
The fingerprint $\mathcal{F} = \{\hat{\boldsymbol{\mu}}_1, \ldots, \hat{\boldsymbol{\mu}}_K\}$ is a set of centroid estimates, one per query, that characterize $p_{\mathcal{P}}(\cdot \mid q_k)$. Averaging over $N$ samples reduces the variance introduced by stochastic decoding (temperature $> 0$). Section~\ref{sec:query_selection} shows that $K = 25$ queries with $N = 3$ samples per query are sufficient for our experiments.

\paragraph{Detection.} Given a suspected deployment using prompt $\mathcal{P}^*$ on model $\mathcal{M}'$, the defender estimates suspect centroids $\hat{\boldsymbol{\mu}}_k^*$ in the same manner and computes the query-level similarity:
\begin{equation}
\sclone = \cos(\hat{\boldsymbol{\mu}}_k, \, \hat{\boldsymbol{\mu}}_k^*).
\label{eq:sim}
\end{equation}
Raw similarity is confounded by query content: for any factual question, outputs from different prompts share substantial semantic overlap simply because they answer the same question. To control for this, the defender also collects outputs under an unrelated prompt $\mathcal{P}_u$ from the same domain and computes $\sunrel = \cos(\hat{\boldsymbol{\mu}}_k, \hat{\boldsymbol{\mu}}_k^{\text{unrel}})$. The detection statistic is the \emph{controlled gap}:
\begin{equation}
\delta_k = \sclone - \sunrel.
\label{eq:gap}
\end{equation}
Under the null hypothesis $H_0$ (the suspect uses a prompt unrelated to $\mathcal{P}$), the expected gap $\mathbb{E}[\delta_k] \approx 0$. Under the alternative $H_1$ (the suspect uses a clone of $\mathcal{P}$), $\mathbb{E}[\delta_k] > 0$ because functional equivalence implies $\sclone > \sunrel$.

\paragraph{Aggregation.} Individual queries provide noisy evidence. For evaluation, we report AUC over all $K$ queries, i.e., $\Pr(\delta_k^{\text{clone}} > \delta_j^{\text{unrel}})$ for randomly drawn $k, j$. For deployment, the decision rule aggregates via a \emph{majority vote}: send $T$ queries, compute $\hat{r} = \frac{1}{T} \sum_{k=1}^{T} \mathbf{1}[\delta_k > 0]$, and declare cloning if $\hat{r} > 0.5$. The proposition below shows that this aggregation yields exponentially increasing confidence as the query budget grows.

\begin{proposition}[Detection guarantee]
\label{prop:guarantee}
Let $p = \Pr(\delta_k > 0 \mid H_1) > 1/2$ be the per-query detection probability under the cloning hypothesis, and assume queries are independent. The majority vote detector over $T$ queries achieves scenario-level detection probability
\begin{equation}
P_{\mathrm{detect}}(T, p) \geq 1 - \exp\big({-2T(p - \tfrac{1}{2})^2}\big).
\label{eq:bound}
\end{equation}
\end{proposition}

A short Hoeffding-style proof is in Appendix~\ref{app:proofs}. For $p = 0.786$ (the empirical query-level recall) and $T = 25$, this bound gives $P_{\mathrm{detect}} \geq 1 - e^{-4.08} > 0.983$. The exact binomial calculation yields $P_{\mathrm{detect}} > 0.999$, and this matches the scenario-level detection probability we observe empirically in non-failure regimes (Section~\ref{sec:deployment}), so the bound, while derived under independence, is not noticeably tight in practice for our query budget. The independence assumption is an approximation: queries from the same benchmark exhibit some positive correlation, which would make a tighter bound depend on the correlation structure. Nevertheless, exponential convergence in $T$ keeps detection reliable under moderate correlation given a sufficient query budget. The bound is informative when $p$ stays well above 0.5; in failure-mode scenarios (Section~\ref{sec:failures}) where $p$ approaches 0.5, reaching equivalent detection confidence calls for a query budget considerably larger than the $T = 25$ default used in our main experiments.

\subsection{Diagnostic Query Optimization}
\label{sec:dqo}

Not all queries carry equal diagnostic value. For a query $q_k$ with a canonical answer, $p_{\mathcal{P}}(\cdot \mid q_k) \approx p_{\mathcal{P}_u}(\cdot \mid q_k)$, so $\delta_k \approx 0$ regardless of whether cloning has occurred. Including such low-information queries in $\mathcal{Q}$ dilutes detection power. Diagnostic Query Optimization (DQO) addresses this by selecting queries that maximize \emph{prompt sensitivity} prior to any interaction with the suspected clone, with empirical validation of the selection rule presented in Section~\ref{sec:query_selection} below.

Define the prompt sensitivity of query $q_k$ as the divergence between the original and unrelated behavioral distributions. Since we do not have access to these distributions directly, DQO uses a tractable proxy: the baseline similarity $\sunrel = \cos(\hat{\boldsymbol{\mu}}_k, \hat{\boldsymbol{\mu}}_k^{\text{unrel}})$. Low $\sunrel$ indicates that the original and unrelated prompts produce highly divergent outputs for $q_k$ (high prompt sensitivity); high $\sunrel$ indicates convergent outputs where prompt variation has little effect. The optimized diagnostic set is:
\begin{equation}
\mathcal{Q}^* = \operatorname*{arg\,min}_{\mathcal{S} \subset \mathcal{Q}_{\text{cand}},\, |\mathcal{S}| = T} \; \sum_{q_k \in \mathcal{S}} \sunrel,
\label{eq:dqo}
\end{equation}
DQO thus reduces to picking, from the candidate pool, the $T$ queries with the smallest baseline similarity to the unrelated prompt. Algorithmically, DQO is a one-shot ranking heuristic rather than a learned model; its contribution is operational (it is \emph{prompt-free} and \emph{transferable}, with the same ranked set reusable across suspected deployments) rather than methodological.

The following proposition provides a sufficient condition under which DQO is optimal.

\begin{proposition}[DQO optimality]
\label{prop:dqo}
If the expected gap $\mathbb{E}[\delta_k \mid H_1]$ is a monotonically decreasing function of baseline similarity $\sunrel$, then the DQO selection rule (Eq.~\ref{eq:dqo}) maximizes the expected total detection signal $\sum_{q_k \in \mathcal{Q}^*} \mathbb{E}[\delta_k \mid H_1]$ over all subsets of size $T$.
\end{proposition}

A short greedy-argument proof is in Appendix~\ref{app:proofs}. The monotonicity assumption is empirically supported: across 32,000 query-scenario pairs, the Pearson correlation between $\sunrel$ and $\delta_k$ is $r = -0.516$ ($p \approx 0$; Section~\ref{sec:query_selection}). This scoring function is also \emph{prompt-free} (it depends only on the defender's own prompts, requiring no access to the suspected clone) and \emph{transferable} (the same ranked query set can be reused across different suspected deployments). Validation on held-out data confirms its effectiveness: the selection rule learned on 125 MMLU training queries lifts AUC from 0.878 to 0.980 on 125 held-out queries (Section~\ref{sec:query_selection}); Algorithm~\ref{alg:bbf} in Appendix~\ref{app:algorithm} gives the full BBF procedure.

\section{Experimental Setup}
\label{sec:setup}

Our evaluation spans four model families and eight benchmarks across 288,000 generated responses.

\paragraph{Models.} The four model families are Claude (claude-haiku-4-5, Anthropic), GPT (gpt-4o-mini, OpenAI), DeepSeek (DeepSeek-V3, via SiliconFlow), and Qwen (qwen-plus, Alibaba). These span four widely-deployed commercial APIs from distinct providers, including both closed-source and open-weight model families.

\paragraph{Benchmarks.} We evaluate on 8 benchmarks spanning 7 domains (Table~\ref{tab:benchmarks}). Diagnostic queries are drawn directly from standard NLP datasets without adversarial crafting or domain-specific engineering. For each benchmark, we design three system prompts: an \emph{original} (the protected prompt), a \emph{clone} (semantic paraphrase preserving functionality), and an \emph{unrelated} (same broad domain, different task). With 4 models forming a $4 \times 4$ cross-model matrix, this yields 16 scenarios per benchmark and 128 total scenarios.

\begin{table}[t]
\centering
\small
\begin{tabular}{llr}
\toprule
\textbf{Domain} & \textbf{Benchmark} & \textbf{Queries} \\
\midrule
\multirow{2}{*}{General QA} & MMLU (57 subjects) & 250 \\
 & TriviaQA & 250 \\
Classification & SST-2 (sentiment) & 250 \\
NLI & MNLI (inference) & 250 \\
Medical & MedQA & 250 \\
Legal & CUAD (contracts) & 250 \\
Math & GSM8K & 250 \\
Summarization & CNN/DailyMail & 250 \\
\midrule
\textbf{Total} & & \textbf{2,000} \\
\bottomrule
\end{tabular}
\caption{Benchmarks used for evaluation. Each benchmark uses 250 queries sampled from standard NLP datasets.}
\label{tab:benchmarks}
\end{table}

\paragraph{Sampling and evaluation.} We use $N = 3$ samples per query, temperature 0.7, and max 400 output tokens. Each experiment generates $250 \times 3 \times 3 = 2{,}250$ model responses (original, clone, and unrelated prompts $\times$ 250 queries $\times$ 3 samples), totaling 288,000 responses across all 128 scenarios. For each query we compare clone similarity to the unrelated baseline, and report AUC over all queries.

\paragraph{Embedding model.} All embeddings use all-MiniLM-L6-v2 \citep{reimers2019sentencebert} (384 dimensions, 22M parameters). Ablation across three architectures (MiniLM-L6, BGE-large, GTE-large) confirms a mean AUC difference below 0.02, as detailed in Section~\ref{sec:embedding_ablation}.

\paragraph{Unrelated prompt design.} For each benchmark, the unrelated prompt shares the same broad domain as the original (e.g., both are medical) but specifies a different task (e.g., diagnosis vs.\ drug interaction). This same-domain design is more conservative than a cross-domain baseline, setting a higher bar for detection by controlling for domain-level similarity.

\section{The Behavioral Imprint of System Prompts}
\label{sec:eta}

\begin{figure}[t]
\centering
\includegraphics[width=\columnwidth]{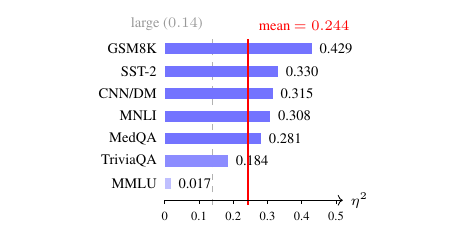}
\caption{Within-benchmark $\eta^2$ effect sizes (CUAD excluded; see Section~\ref{sec:failures}). Dashed line: Cohen's large-effect threshold ($\eta^2 > 0.14$).}
\label{fig:eta}
\end{figure}

Section~\ref{sec:intro} previewed our central observation: system prompt choice has a large, measurable effect on model outputs, and this is what licenses output similarity as a detection signal. We quantify that effect via a one-way ANOVA in which prompt identity (original / clone / unrelated) is the factor and output-embedding similarity is the dependent variable, computing $\eta^2$ within each benchmark to avoid cross-benchmark variance inflation (Figure~\ref{fig:eta}).

The mean $\eta^2 = 0.244$ constitutes a large effect by Cohen's standard \citep{cohen1988}, whose thresholds for $\eta^2$ are 0.01 (small), 0.06 (medium), and 0.14 (large); 0.244 is roughly 1.7$\times$ the large-effect threshold and means that prompt choice explains on average 24.4\% of within-benchmark output variance. The effect varies substantially across domains: GSM8K shows the strongest prompt influence ($\eta^2 = 0.429$), consistent with the diversity of step-by-step reasoning styles that different prompts elicit, while MMLU shows a small effect ($\eta^2 = 0.017$), reflecting its many canonically-answered questions where any prompt produces similar responses. CUAD is excluded here due to a HuggingFace dataset issue (Section~\ref{sec:failures}) that introduced a prompt-query domain mismatch; AUC-based analyses elsewhere in the paper are unaffected by this artifact.

This establishes prompt as a load-bearing factor in output variance, a necessary but not sufficient condition for detection; whether the signal actually separates clone from unrelated outputs is the empirical question Section~\ref{sec:results} addresses next.

\section{Main Results}
\label{sec:results}

Having established that system prompts leave measurable behavioral fingerprints, we now evaluate whether this signal is sufficient for practical clone detection. We report results across all 128 scenarios, organized by same-model detection, cross-model transfer, and failure modes.

\subsection{Overview}
\label{sec:overview}

Table~\ref{tab:auc_summary} summarizes all 128 scenarios: overall mean AUC 0.763 and same-model 0.876. Cross-model performance is governed by which model serves as detector: off-diagonal column means span 0.845 (Claude) to 0.665 (Qwen), with overall cross-model mean 0.725 (Section~\ref{sec:crossmodel} and Table~\ref{tab:matrix}). The 11 failures (AUC $\leq 0.50$) concentrate in CUAD (5), SST-2 (3), and CNN/DM (3); excluding CUAD's data-loading issue (Section~\ref{sec:failures}), the method-attributable failure rate is 6 of 112 (5.4\%).

\begin{table}[t]
\centering
\small
\setlength{\tabcolsep}{4pt}
\begin{tabular}{lrrrrr}
\toprule
\textbf{Benchmark} & $n$ & \textbf{Mean} & \textbf{Min} & \textbf{Max} & \textbf{Fails} \\
\midrule
MMLU & 16 & 0.720 & 0.500 & 0.918 & 0 \\
TriviaQA & 16 & 0.796 & 0.676 & 0.944 & 0 \\
SST-2 & 16 & 0.775 & 0.239 & 1.000 & 3 \\
MNLI & 16 & 0.823 & 0.589 & 0.983 & 0 \\
MedQA & 16 & 0.798 & 0.529 & 0.999 & 0 \\
CUAD$^\dagger$ & 16 & 0.651 & 0.235 & 0.987 & 5 \\
GSM8K & 16 & 0.764 & 0.544 & 0.977 & 0 \\
CNN/DM & 16 & 0.773 & 0.449 & 1.000 & 3 \\
\midrule
\textbf{All} & 128 & 0.763 & 0.235 & 1.000 & 11 \\
\textbf{Excl.\ CUAD} & 112 & 0.779 & 0.239 & 1.000 & 6 \\
\bottomrule
\end{tabular}
\caption{Per-benchmark AUC summary. $n$ = scenarios; ``Fails'' = AUC $\leq 0.50$. $^\dagger$CUAD results are affected by a dataset loading issue (see Section~\ref{sec:failures}); the ``Excl.\ CUAD'' row provides method-only failure rates (6/112 = 5.4\%).}
\label{tab:auc_summary}
\end{table}

\begin{table}[t]
\centering
\small
\setlength{\tabcolsep}{4pt}
\begin{tabular}{lcccc|c}
\toprule
\multirow{2}{*}{\textbf{Reg.}} & \multicolumn{4}{c|}{\textbf{Detection Model}} & \multirow{2}{*}{\textbf{Mean}} \\
 & \textbf{Claude} & \textbf{GPT} & \textbf{DS} & \textbf{Qwen} & \\
\midrule
Claude & \textbf{.967} & .693 & .688 & .619 & .742 \\
GPT & .830 & \textbf{.875} & .719 & .736 & .790 \\
DeepSeek & .863 & .782 & \textbf{.867} & .640 & .788 \\
Qwen & .843 & .649 & .636 & \textbf{.796} & .731 \\
\midrule
\textbf{Col.\ mean} & \textbf{.876} & .750 & .728 & .698 & .763 \\
\bottomrule
\end{tabular}
\caption{Mean AUC across 8 benchmarks for each registration $\times$ detection model pair. Bold diagonal = same-model. Column means reveal that \emph{detection} model identity matters most ($R^2 = 0.161$), exceeding benchmark ($R^2 = 0.096$) and registration model ($R^2 = 0.050$).}
\label{tab:matrix}
\end{table}

\subsection{Same-Model Detection}
\label{sec:samemodel}

When the fingerprint is registered and the suspect runs on the same model family, BBF operates in its most common detection setting. Same-model detection is reliable, with an overall mean AUC of 0.876 across 32 scenarios (per-model means: Claude 0.967, GPT 0.875, DeepSeek 0.867, Qwen 0.796). Claude stands out for its consistency (6 of 8 benchmarks exceed 0.94). Only one same-model scenario fails (CUAD on GPT$\to$GPT, 0.448), discussed in Section~\ref{sec:failures}. The full $4 \times 8$ benchmark-by-model breakdown is in Appendix~\ref{app:samemodel} (Table~\ref{tab:samemodel_full}).

\subsection{Cross-Model Detection}
\label{sec:crossmodel}

In practice, an adversary deploys the stolen prompt on a different model. Restricted to the 12 off-diagonal cross-model pairs in Table~\ref{tab:matrix}, mean AUC ranges from 0.845 with Claude as detector down to 0.665 with Qwen. Detector-model identity accounts for $R^2 = 0.161$ of AUC variance across the full matrix, dominating the $R^2 = 0.050$ contributed by registration choice.

Transfer works because prompt-specific patterns (reasoning steps, caveats, structure) are partially model-agnostic and captured by sentence embeddings. Gaps are sometimes \emph{larger} cross-model than same-model (e.g., Claude$\to$GPT +0.101 vs.\ Claude$\to$Claude +0.093 on MMLU), because unrelated prompts diverge more across families.

\subsection{Failure Modes}
\label{sec:failures}

Of 128 scenarios, 11 produce AUC $\leq 0.50$. The failures cluster into three distinct patterns:

\paragraph{Short-output directional failures (SST-2, 3 failures).} Claude-registered fingerprints fail to transfer outward (C$\to$G = 0.239, C$\to$D = 0.264), and Qwen$\to$GPT fails (0.495). Sentiment outputs are extremely short (often $<$50 tokens), and Claude produces idiosyncratic concise formats. The failure is directional: GPT and DeepSeek fingerprints transfer effectively. Output above 50 tokens is therefore a necessary precondition for reliable transfer across model families.

\paragraph{Data artifact failures (CUAD, 5 failures).} CUAD contributes 5 failures including the sole same-model failure (GG = 0.448). These are \emph{data artifacts}, not method failures: the HuggingFace cuad loader silently fell back to a SQuAD-formatted split when the expected contract-clause configuration was unavailable, so our contract-analysis prompts were issued against reading-comprehension passages. The mismatch suppresses the prompt-conditional signal because outputs collapse to passage-grounded extraction regardless of system prompt. The remaining seven benchmarks load from standard, stable splits (MMLU, TriviaQA, SST-2, MNLI, MedQA, GSM8K, CNN/DM) and exhibit no analogous prompt-query domain mismatch in the queries we sampled. Excluding CUAD, the method-attributable failure rate drops to 6 of 112 scenarios, or 5.4\%.

\paragraph{Cross-model style mismatch (CNN/DM, 3 failures).} DeepSeek$\to$Qwen (0.500), Qwen$\to$GPT (0.449), and Qwen$\to$DeepSeek (0.474) reflect a systematic style mismatch between these pairs (also seen in MMLU D$\to$Q = 0.500). Both models transfer well to Claude (D$\to$C, Q$\to$C $> 0.85$), confirming the failure is pair-specific.

\section{Ablation Studies}
\label{sec:ablation}

Section~\ref{sec:query_selection} ablates the query selection rule and the query budget $K$, and Section~\ref{sec:embedding_ablation} compares the embedding model against alternative encoders and against lexical similarity baselines.

\subsection{Query Discriminability and DQO Validation}
\label{sec:query_selection}

\begin{figure*}[t]
\centering
\includegraphics[width=\textwidth]{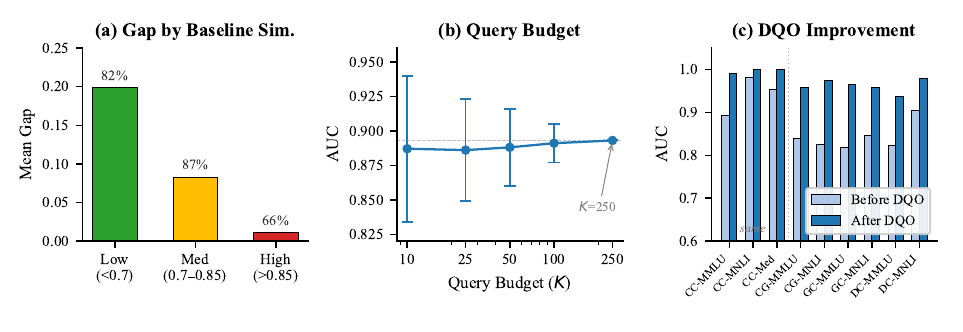}
\caption{Query selection analysis. \emph{(a)} Detection gap by baseline-similarity bin; percentages mark the proportion of positive gaps. \emph{(b)} Query budget ablation on MMLU Claude$\to$Claude. \emph{(c)} DQO before/after across 9 settings.}
\label{fig:queries}
\end{figure*}

Some queries yield non-positive gaps because their content is canonical enough that any prompt produces similar output (Pearson $r = -0.516$ between $\sunrel$ and $\delta_k$, $p \approx 0$, $n = 32{,}000$). As Figure~\ref{fig:queries}a shows, low-baseline queries yield the largest gaps (mean +0.199 for $s^{\text{unrel}} < 0.7$ vs.\ +0.011 for $s^{\text{unrel}} > 0.85$), motivating DQO (Section~\ref{sec:dqo}). DQO improves AUC in all 9 validated settings (Figure~\ref{fig:queries}c), with larger gains in cross-model scenarios. A query budget of $K = 25$ already suffices for reliable detection (Figure~\ref{fig:queries}b).

\begin{table}[t]
\centering
\small
\begin{tabular}{lcccc}
\toprule
\textbf{Bench.} & \textbf{Orig.} & \textbf{L1} & \textbf{L2} & \textbf{L3} \\
\midrule
MMLU & .862 & .966 & .922 & .889 \\
SST-2 & .996 & .984 & 1.00 & .969 \\
MNLI & .978 & .997 & .984 & .991 \\
MedQA & .933 & .983 & .958 & .956 \\
GSM8K & .991 & .997 & .998 & 1.00 \\
CNN/DM & 1.00 & 1.00 & 1.00 & 1.00 \\
\bottomrule
\end{tabular}
\caption{Paraphrase attack (Claude$\to$Claude): even under L3, all AUC $\geq 0.889$. The companion table on LLM-simulated extraction quality is in Appendix~\ref{app:extraction} (Table~\ref{tab:extraction}).}
\label{tab:paraphrase}
\end{table}

\subsection{Embedding Model and Baseline Comparison}
\label{sec:embedding_ablation}

Detection is consistent across embedding backends: mean AUC difference is below 0.02 across MiniLM-L6, BGE-large, and GTE-large. Lexical baselines (TF-IDF, bigram Jaccard) are competitive on same-model pairs but degrade on cross-model pairs (e.g., MNLI Claude$\to$GPT: BBF 0.893 vs.\ TF-IDF 0.772), confirming BBF's advantage where surface similarity breaks down. Full per-benchmark numbers are in Appendix~\ref{app:embedding} (Table~\ref{tab:embedding_full}).

\section{Robustness}
\label{sec:robustness}

Section~\ref{sec:paraphrase} evaluates paraphrase attacks; Section~\ref{sec:adaptive} evaluates adaptive prefix injection; Section~\ref{sec:deployment} covers practical deployment conditions.

\subsection{Non-Adaptive Paraphrase Attack}
\label{sec:paraphrase}

GPT-4o-mini generates rewrites at three intensity levels: L1 (synonym substitution), L2 (structural reorganization), and L3 (complete functional rewrite), evaluated on Claude$\to$Claude across 6 benchmarks (50 queries each). Even under L3, all AUC values remain $\geq 0.889$ (Table~\ref{tab:paraphrase}). L1 paraphrase sometimes raises AUC above the original clone: MMLU rises from 0.862 to 0.966, suggesting that light rewrites can smooth stylistic noise.

\begin{table}[t]
\centering
\small
\begin{tabular}{lcc}
\toprule
\textbf{Prefix Strategy} & \textbf{MMLU} & \textbf{MNLI} \\
\midrule
None (baseline) & 0.855 & 0.978 \\
Chain-of-thought & 0.792 & 0.958 \\
Formal academic tone & 0.863 & 0.547 \\
Random padding & 0.838 & 0.771 \\
\bottomrule
\end{tabular}
\caption{Adaptive prefix injection (Claude$\to$Claude, 50 queries). Formal tone evades MNLI detection (0.978 $\to$ 0.547) but barely shifts MMLU, which holds at 0.863.}
\label{tab:prefix}
\end{table}

\subsection{Adaptive Attack: Prefix Injection}
\label{sec:adaptive}

Paraphrasing the prompt fails to evade BBF (\S\ref{sec:paraphrase}), because the embedding signature tracks behavior rather than wording. An adversary who instead targets the \emph{output style}, by prepending a short instruction that overrides the prompt's stylistic surface, poses a stronger threat. We evaluate four such prefix-injection strategies on MMLU and MNLI (Claude$\to$Claude, 50 queries).

Table~\ref{tab:prefix} reveals an asymmetry: Non-adaptive paraphrasing does not evade BBF, but targeted style-shifting can degrade detection on short, structured outputs (MNLI: 0.978 $\to$ 0.547 under formal tone) while leaving content-rich outputs unaffected (MMLU: 0.863). Defending against style-shifting on short outputs remains open.

\subsection{Deployment Conditions}
\label{sec:deployment}

We evaluate three practical deployment factors. \textbf{Imperfect extraction}: On MMLU (Claude$\to$Claude), author-constructed clones at all five fidelity levels achieve AUC $\geq 0.879$, and LLM-simulated clones match or exceed the handcrafted baseline (AUC 0.886 $\to$ 0.982; Appendix Table~\ref{tab:extraction}). The direction of this result is counter-intuitive: \emph{noisier} extracted clones are at least as detectable as carefully handcrafted paraphrases. Our hypothesis is that extracted text preserves the original prompt's functional framing and surface phrasing even when imperfect, whereas a handcrafted paraphrase, written by an independent author with their own style, injects orthogonal stylistic variance that the embedding picks up as noise. The practical implication is favorable: the realistic threat model, in which an adversary deploys text obtained from automated extraction \citep{hui2024pleak, yao2025prsa} rather than rewriting the prompt from scratch, is precisely the regime where BBF is strongest. \textbf{Cross-domain generalization}: LOOCV across the 8 NLP benchmarks yields mean AUC = 0.793, with CUAD remaining the lowest at 0.651; a supplementary check on four enterprise domains (legal, financial, clinical, software architecture) confirms positive signal (mean gap +0.194, 15/16 significant; Appendix~\ref{app:enterprise}). \textbf{Detection thresholds}: query-level recall at gap $> 0$ is 78.6\% across 32,000 pairs; majority vote over $T = 25$ then lifts the scenario-level detection probability above 0.999.

\section{Conclusion}
\label{sec:conclusion}

BBF detects cloned system prompts from black-box outputs at AUC 0.876 same-model and 0.725 mean cross-model, with detector identity (0.845 Claude to 0.665 Qwen) as the dominant transfer factor. The signal is reliable for content-rich outputs and Claude or DeepSeek detectors, but collapses under output-style adaptation on short structured tasks (Section~\ref{sec:adaptive}); closing that style-invariance gap is the key direction this work leaves open. Within these limits, BBF gives a prompt owner a defensible verification protocol that requires only black-box API access, no platform cooperation, and a fixed budget of 25 diagnostic queries.

\section*{Limitations}

The strongest open issue is robustness to \emph{adaptive} attacks. Section~\ref{sec:adaptive} shows that a single-sentence formal-tone prefix collapses MNLI detection from 0.978 to 0.547, and our evaluation covers only four prefix-injection strategies on two benchmarks; sandwich constructions, suffix steering, output-format steering, and multi-turn dilution remain unexplored. Style-invariant detection on short, structured outputs is therefore the natural next direction, and BBF should not be deployed against adversaries assumed to be aware of the detector. A second limitation is scope: BBF is evaluated on 8 text-based NLP benchmarks and four mid-tier commercial APIs, so generalization to multimodal tasks, real-time agents, open-source or frontier-scale models, and sandwich/RAG deployments is not established. Our evaluation is also single-turn and does not cover the agent / tool-use settings motivated in Section~\ref{sec:intro}. Finally, LOOCV gives feasible but imperfect cross-domain transfer (mean AUC = 0.793), so domain-specific calibration may be needed.

\section*{Ethics Statement}

BBF provides statistical evidence of prompt cloning, not legal conclusions; the legal status of system prompt cloning varies by jurisdiction. We recommend it be used only with reasonable prior suspicion of cloning, not for speculative surveillance. Diagnostic queries are drawn from established public NLP benchmarks; code and data will be released with the camera-ready version.


\appendix

\section{Algorithm and Proofs}
\label{app:algorithm}

\subsection{Algorithm}

\begin{algorithm}[h]
\caption{Black-Box Behavioral Fingerprinting}
\label{alg:bbf}
\begin{algorithmic}[1]
\REQUIRE Original prompt $\mathcal{P}$, unrelated prompt $\mathcal{P}_u$, candidate queries $\mathcal{Q}_{\text{cand}}$, budget $T$, samples $N$, encoder $\phi$
\ENSURE Decision: \textsc{Clone} or \textsc{Not Clone}
\STATE \emph{// Phase 1: Registration \& Query Optimization}
\FOR{$q_k \in \mathcal{Q}_{\text{cand}}$}
    \STATE $\hat{\boldsymbol{\mu}}_k \gets \frac{1}{N}\sum_{i=1}^{N} \phi(\mathcal{M}(q_k; \mathcal{P})_i)$
    \STATE $\hat{\boldsymbol{\mu}}_k^{\text{unrel}} \gets \frac{1}{N}\sum_{i=1}^{N} \phi(\mathcal{M}(q_k; \mathcal{P}_u)_i)$
    \STATE $\sunrel \gets \cos(\hat{\boldsymbol{\mu}}_k, \hat{\boldsymbol{\mu}}_k^{\text{unrel}})$
\ENDFOR
\STATE $\mathcal{Q}^* \gets$ top-$T$ queries with lowest $\sunrel$ \hfill \emph{// DQO}
\STATE \emph{// Phase 2: Detection}
\FOR{$q_k \in \mathcal{Q}^*$}
    \STATE $\hat{\boldsymbol{\mu}}_k^* \gets \frac{1}{N}\sum_{i=1}^{N} \phi(\mathcal{M}'(q_k; \mathcal{P}^*)_i)$
    \STATE $\delta_k \gets \cos(\hat{\boldsymbol{\mu}}_k, \hat{\boldsymbol{\mu}}_k^*) - \sunrel$ \hfill \emph{// from Phase 1}
\ENDFOR
\STATE $\hat{r} \gets \frac{1}{T}\sum_{k=1}^{T} \mathbf{1}[\delta_k > 0]$
\IF{$\hat{r} > 0.5$}
    \RETURN \textsc{Clone}
\ELSE
    \RETURN \textsc{Not Clone}
\ENDIF
\end{algorithmic}
\end{algorithm}

\subsection{Proofs}
\label{app:proofs}

\paragraph{Proposition 1 (Detection guarantee).} The number of positive-gap queries $S = \sum_{k=1}^{T} \mathbf{1}[\delta_k > 0]$ follows $\text{Binomial}(T, p)$ with $\mathbb{E}[S] = Tp$. A majority vote fails when $S \leq T/2$, i.e., $S - Tp \leq T(1/2 - p)$. By Hoeffding's inequality, $\Pr(S \leq T/2) \leq \exp(-2T(p - 1/2)^2)$. \hfill $\square$

\paragraph{Proposition 2 (DQO optimality).} Under the monotonicity assumption, ordering queries by $\sunrel$ in ascending order is equivalent to ordering by $\mathbb{E}[\delta_k \mid H_1]$ in descending order. Selecting the top-$T$ by descending expected gap maximizes $\sum_{q_k \in \mathcal{S}} \mathbb{E}[\delta_k \mid H_1]$ by a standard greedy argument for modular functions. \hfill $\square$

\section{Supplementary Tables and Figures}
\label{app:supplementary}

\subsection{Same-Model Detection: Full Breakdown}
\label{app:samemodel}

\begin{table}[h]
\centering
\small
\begin{tabular}{lcccc}
\toprule
\textbf{Benchmark} & \textbf{CC} & \textbf{GG} & \textbf{DD} & \textbf{QQ} \\
\midrule
MMLU & 0.893 & 0.918 & 0.781 & 0.654 \\
TriviaQA & 0.944 & 0.944 & 0.735 & 0.838 \\
SST-2 & 0.998 & 0.991 & 1.000 & 0.793 \\
MNLI & 0.983 & 0.964 & 0.925 & 0.836 \\
MedQA & 0.953 & 0.959 & 0.999 & 0.671 \\
CUAD & 0.987 & 0.448$^*$ & 0.918 & 0.830 \\
GSM8K & 0.977 & 0.976 & 0.793 & 0.755 \\
CNN/DM & 1.000 & 0.797 & 0.783 & 0.988 \\
\midrule
\textbf{Mean} & \textbf{0.967} & \textbf{0.875} & \textbf{0.867} & \textbf{0.796} \\
\bottomrule
\end{tabular}
\caption{Same-model AUC. CC = Claude$\to$Claude, GG = GPT$\to$GPT, DD = DeepSeek$\to$DeepSeek, QQ = Qwen$\to$Qwen. $^*$CUAD GG is the only same-model failure (0.448). Overall same-model mean = 0.876.}
\label{tab:samemodel_full}
\end{table}

\subsection{Embedding Model and Lexical Baseline Ablation}
\label{app:embedding}

We compare MiniLM-L6 against two larger embedding models (BGE-large, GTE-large; both 1024d) and two lexical baselines (TF-IDF, bigram Jaccard).

\begin{table}[h]
\centering
\small
\setlength{\tabcolsep}{4pt}
\begin{tabular}{llcccc}
\toprule
 & & \multicolumn{3}{c}{\textbf{Embeddings}} & \textbf{Lexical} \\
\cmidrule(lr){3-5} \cmidrule(lr){6-6}
\textbf{Bench.} & \textbf{Pair} & \textbf{MiniLM} & \textbf{BGE} & \textbf{GTE} & \textbf{TF-IDF} \\
\midrule
\multirow{2}{*}{MMLU} & CC & .891 & .945 & .918 & .927 \\
 & CG & .833 & --- & --- & .794 \\
SST-2 & CC & .997 & .996 & .998 & --- \\
\multirow{2}{*}{MNLI} & CC & .987 & .988 & .983 & .930 \\
 & CG & .893 & --- & --- & .772 \\
MedQA & CC & .940 & .937 & .946 & --- \\
GSM8K & CC & .998 & .998 & .994 & --- \\
CNN/DM & CC & 1.00 & 1.00 & 1.00 & --- \\
\bottomrule
\end{tabular}
\caption{Similarity measure comparison. Embedding models: mean AUC difference $< 0.02$. TF-IDF: competitive on same-model (CC) but degrades on cross-model (CG) pairs.}
\label{tab:embedding_full}
\end{table}

\subsection{Extraction Quality Robustness}
\label{app:extraction}

LLM-simulated extracted clones at four fidelity levels, on MMLU and MNLI (Claude$\to$Claude), compared to handcrafted paraphrases.

\begin{table}[h]
\centering
\small
\setlength{\tabcolsep}{4pt}
\begin{tabular}{llccc}
\toprule
\textbf{Bench.} & \textbf{Clone type} & \textbf{AUC} & \textbf{Gap} & $\bar{s}_{\text{cl.}}$ \\
\midrule
\multirow{4}{*}{MMLU} & Handcrafted & .886 & +.091 & .937 \\
 & Extr.\ (good) & .935 & +.110 & .951 \\
 & Extr.\ (partial) & .959 & +.109 & .954 \\
 & Extr.\ (noisy) & .982 & +.121 & .963 \\
\midrule
\multirow{4}{*}{MNLI} & Handcrafted & .985 & +.092 & .949 \\
 & Extr.\ (good) & .984 & +.102 & .953 \\
 & Extr.\ (partial) & .980 & +.096 & .959 \\
 & Extr.\ (noisy) & 1.00 & +.119 & .971 \\
\bottomrule
\end{tabular}
\caption{Extraction quality (Claude$\to$Claude). LLM-simulated extracted clones are at least as detectable as handcrafted paraphrases; noisier extractions are slightly more detectable, consistent with the hypothesis discussed in Section~\ref{sec:deployment}.}
\label{tab:extraction}
\end{table}

\subsection{Cross-Model AUC Heatmaps}
\label{app:heatmaps}

\begin{figure*}[t]
\centering
\includegraphics[width=\textwidth]{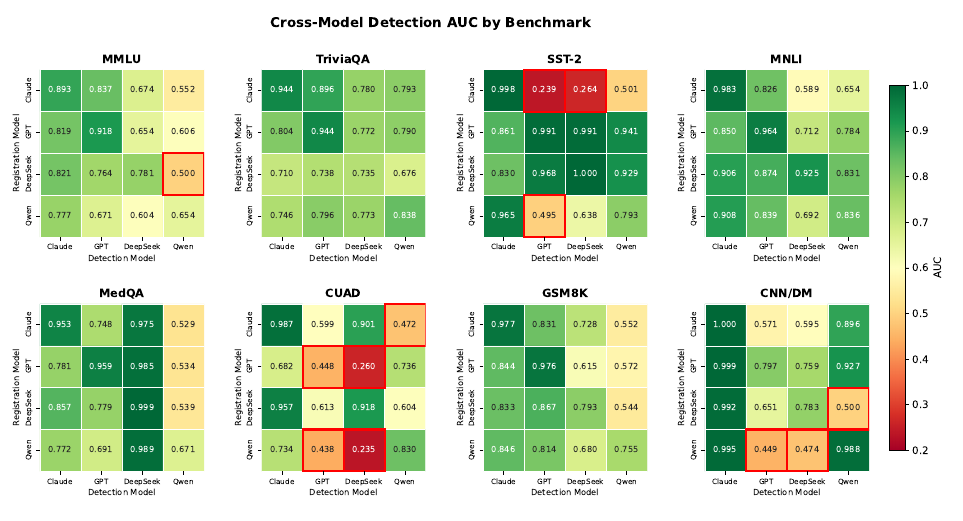}
\caption{Cross-model detection AUC heatmaps for all 8 benchmarks. Each cell shows the AUC when registering a fingerprint on the row model and detecting on the column model. Red borders mark failures (AUC $\leq 0.50$). Failures concentrate in CUAD (5), SST-2 (3), and CNN/DM (3), primarily involving Claude-registered fingerprints detected on GPT/DeepSeek (short-output tasks) or GPT-registered fingerprints on DeepSeek (CUAD).}
\label{fig:heatmaps}
\end{figure*}

Figure~\ref{fig:heatmaps} visualizes the $4 \times 4$ cross-model AUC matrices for each benchmark as heatmaps, with red-bordered cells marking detection failures where AUC drops to 0.50 or below.

\section{Enterprise Skill Domain Validation}
\label{app:enterprise}

To probe generalization beyond NLP benchmarks, we evaluate BBF on four enterprise system-prompt domains: legal contract analysis, equity valuation, clinical differential diagnosis, and software architecture review. Each domain uses 5 boundary queries hand-designed to maximize prompt sensitivity, with $N = 3$ samples per query, evaluated across the same $2 \times 2$ Claude/GPT same/cross-model design (16 scenarios). Across all 16, mean controlled gap is +0.194 (range +0.034 to +0.646), with 15 of 16 significant at $p < 0.05$. Same-model mean AUC is 0.96 and cross-model is 0.92; the small per-domain query budget inflates AUC ceilings, so we report gap as the primary statistic. The single non-significant scenario is software architecture under Claude$\to$GPT transfer ($p = 0.21$), echoing the main-result observation that programming-style outputs vary more across model families. These results indicate that BBF generalizes beyond standardized NLP benchmarks to enterprise system-prompt deployments, though sample sizes warrant a larger follow-up study.

\section{System Prompts and Full Per-Scenario Results}
\label{app:prompts}

For each benchmark, we design three prompts: an \emph{original} (the protected system prompt), a \emph{clone} (semantic paraphrase preserving functionality), and an \emph{unrelated} (same broad domain but different task). Full prompt texts for all 8 benchmarks, together with per-query similarity scores and gap distributions for the 128 scenarios, are provided in the supplementary materials. Table~\ref{tab:all_scenarios} lists the 128 scenario identifiers and their AUC values.

\subsection{Example Prompt Triplet (MedQA)}
\label{app:medqa_triplet}

To make the design concrete, we reproduce the full three-prompt set used for the MedQA benchmark. The \emph{original} and \emph{clone} target the same skill (board-style clinical reasoning) with different surface phrasing and structure; the \emph{unrelated} prompt stays in the medical domain but specifies a different task (curriculum design), so it controls for domain-level overlap without sharing the protected prompt's functionality.

\paragraph{Original.}
\begin{quote}\small\ttfamily
You are a clinical reasoning assistant for medical board-style questions. When analyzing medical questions: identify the key clinical presentation (symptoms, signs, lab values, imaging); generate a differential diagnosis ranked by likelihood; use pathophysiology to connect findings to diagnoses; for multiple-choice, eliminate options systematically using clinical reasoning; distinguish between similar conditions (e.g., Crohn's vs.\ UC, Type 1 vs.\ Type 2 DM); apply clinical decision rules and guidelines where applicable; note when a question tests a classic ``buzzword'' association. Format: (1) Key findings, (2) Differential, (3) Reasoning, (4) Answer.
\end{quote}

\paragraph{Clone (semantic paraphrase).}
\begin{quote}\small\ttfamily
You are a medical knowledge assistant for board exam preparation. For each clinical question: extract critical clinical data (symptoms, vitals, labs, imaging results); build a ranked differential diagnosis list; link findings to underlying pathophysiology; systematically eliminate wrong answer choices with clinical logic; differentiate look-alike conditions using distinguishing features; apply relevant clinical guidelines and scoring systems; recognize classic presentation patterns and associations. Format: (1) Clinical clues, (2) Differentials, (3) Analysis, (4) Best answer.
\end{quote}

\paragraph{Unrelated (same domain, different task).}
\begin{quote}\small\ttfamily
You are a medical education curriculum designer. Design case-based learning modules for medical students; create standardized patient encounter scripts; develop OSCE station checklists with scoring rubrics; map learning objectives to medical competency frameworks; recommend simulation scenarios for procedural skills training. Format: (1) Learning objectives, (2) Case outline, (3) Assessment rubric.
\end{quote}

\begin{table*}[t]
\centering
\small
\setlength{\tabcolsep}{5pt}
\begin{tabular}{rllc|rllc}
\toprule
\textbf{\#} & \textbf{Bench.} & \textbf{Scenario} & \textbf{AUC} & \textbf{\#} & \textbf{Bench.} & \textbf{Scenario} & \textbf{AUC} \\
\midrule
1 & MMLU & C$\to$C & 0.893 & 65 & MedQA & C$\to$C & 0.953 \\
2 & MMLU & C$\to$G & 0.837 & 66 & MedQA & C$\to$G & 0.748 \\
3 & MMLU & C$\to$D & 0.674 & 67 & MedQA & C$\to$D & 0.975 \\
4 & MMLU & C$\to$Q & 0.552 & 68 & MedQA & C$\to$Q & 0.529 \\
5 & MMLU & G$\to$C & 0.819 & 69 & MedQA & G$\to$C & 0.781 \\
6 & MMLU & G$\to$G & 0.918 & 70 & MedQA & G$\to$G & 0.959 \\
7 & MMLU & G$\to$D & 0.654 & 71 & MedQA & G$\to$D & 0.985 \\
8 & MMLU & G$\to$Q & 0.606 & 72 & MedQA & G$\to$Q & 0.534 \\
9 & MMLU & D$\to$C & 0.821 & 73 & MedQA & D$\to$C & 0.857 \\
10 & MMLU & D$\to$G & 0.764 & 74 & MedQA & D$\to$G & 0.779 \\
11 & MMLU & D$\to$D & 0.781 & 75 & MedQA & D$\to$D & 0.999 \\
12 & MMLU & D$\to$Q & 0.500 & 76 & MedQA & D$\to$Q & 0.539 \\
13 & MMLU & Q$\to$C & 0.777 & 77 & MedQA & Q$\to$C & 0.772 \\
14 & MMLU & Q$\to$G & 0.671 & 78 & MedQA & Q$\to$G & 0.691 \\
15 & MMLU & Q$\to$D & 0.604 & 79 & MedQA & Q$\to$D & 0.989 \\
16 & MMLU & Q$\to$Q & 0.654 & 80 & MedQA & Q$\to$Q & 0.671 \\
\midrule
17 & TriviaQA & C$\to$C & 0.944 & 81 & CUAD & C$\to$C & 0.987 \\
18 & TriviaQA & C$\to$G & 0.896 & 82 & CUAD & C$\to$G & 0.599 \\
19 & TriviaQA & C$\to$D & 0.780 & 83 & CUAD & C$\to$D & 0.901 \\
20 & TriviaQA & C$\to$Q & 0.793 & 84 & CUAD & C$\to$Q & 0.472$^*$ \\
21 & TriviaQA & G$\to$C & 0.804 & 85 & CUAD & G$\to$C & 0.682 \\
22 & TriviaQA & G$\to$G & 0.944 & 86 & CUAD & G$\to$G & 0.448$^*$ \\
23 & TriviaQA & G$\to$D & 0.772 & 87 & CUAD & G$\to$D & 0.260$^*$ \\
24 & TriviaQA & G$\to$Q & 0.790 & 88 & CUAD & G$\to$Q & 0.736 \\
25 & TriviaQA & D$\to$C & 0.710 & 89 & CUAD & D$\to$C & 0.957 \\
26 & TriviaQA & D$\to$G & 0.738 & 90 & CUAD & D$\to$G & 0.613 \\
27 & TriviaQA & D$\to$D & 0.735 & 91 & CUAD & D$\to$D & 0.918 \\
28 & TriviaQA & D$\to$Q & 0.676 & 92 & CUAD & D$\to$Q & 0.604 \\
29 & TriviaQA & Q$\to$C & 0.746 & 93 & CUAD & Q$\to$C & 0.734 \\
30 & TriviaQA & Q$\to$G & 0.796 & 94 & CUAD & Q$\to$G & 0.438$^*$ \\
31 & TriviaQA & Q$\to$D & 0.773 & 95 & CUAD & Q$\to$D & 0.235$^*$ \\
32 & TriviaQA & Q$\to$Q & 0.838 & 96 & CUAD & Q$\to$Q & 0.830 \\
\midrule
33 & SST-2 & C$\to$C & 0.998 & 97 & GSM8K & C$\to$C & 0.977 \\
34 & SST-2 & C$\to$G & 0.239$^*$ & 98 & GSM8K & C$\to$G & 0.831 \\
35 & SST-2 & C$\to$D & 0.264$^*$ & 99 & GSM8K & C$\to$D & 0.728 \\
36 & SST-2 & C$\to$Q & 0.501 & 100 & GSM8K & C$\to$Q & 0.552 \\
37 & SST-2 & G$\to$C & 0.861 & 101 & GSM8K & G$\to$C & 0.844 \\
38 & SST-2 & G$\to$G & 0.991 & 102 & GSM8K & G$\to$G & 0.976 \\
39 & SST-2 & G$\to$D & 0.991 & 103 & GSM8K & G$\to$D & 0.615 \\
40 & SST-2 & G$\to$Q & 0.941 & 104 & GSM8K & G$\to$Q & 0.572 \\
41 & SST-2 & D$\to$C & 0.830 & 105 & GSM8K & D$\to$C & 0.833 \\
42 & SST-2 & D$\to$G & 0.968 & 106 & GSM8K & D$\to$G & 0.867 \\
43 & SST-2 & D$\to$D & 1.000 & 107 & GSM8K & D$\to$D & 0.793 \\
44 & SST-2 & D$\to$Q & 0.929 & 108 & GSM8K & D$\to$Q & 0.544 \\
45 & SST-2 & Q$\to$C & 0.965 & 109 & GSM8K & Q$\to$C & 0.846 \\
46 & SST-2 & Q$\to$G & 0.495$^*$ & 110 & GSM8K & Q$\to$G & 0.814 \\
47 & SST-2 & Q$\to$D & 0.638 & 111 & GSM8K & Q$\to$D & 0.680 \\
48 & SST-2 & Q$\to$Q & 0.793 & 112 & GSM8K & Q$\to$Q & 0.755 \\
\midrule
49 & MNLI & C$\to$C & 0.983 & 113 & CNN/DM & C$\to$C & 1.000 \\
50 & MNLI & C$\to$G & 0.826 & 114 & CNN/DM & C$\to$G & 0.571 \\
51 & MNLI & C$\to$D & 0.589 & 115 & CNN/DM & C$\to$D & 0.595 \\
52 & MNLI & C$\to$Q & 0.654 & 116 & CNN/DM & C$\to$Q & 0.896 \\
53 & MNLI & G$\to$C & 0.850 & 117 & CNN/DM & G$\to$C & 0.999 \\
54 & MNLI & G$\to$G & 0.964 & 118 & CNN/DM & G$\to$G & 0.797 \\
55 & MNLI & G$\to$D & 0.712 & 119 & CNN/DM & G$\to$D & 0.759 \\
56 & MNLI & G$\to$Q & 0.784 & 120 & CNN/DM & G$\to$Q & 0.927 \\
57 & MNLI & D$\to$C & 0.906 & 121 & CNN/DM & D$\to$C & 0.992 \\
58 & MNLI & D$\to$G & 0.874 & 122 & CNN/DM & D$\to$G & 0.651 \\
59 & MNLI & D$\to$D & 0.925 & 123 & CNN/DM & D$\to$D & 0.783 \\
60 & MNLI & D$\to$Q & 0.831 & 124 & CNN/DM & D$\to$Q & 0.500$^*$ \\
61 & MNLI & Q$\to$C & 0.908 & 125 & CNN/DM & Q$\to$C & 0.995 \\
62 & MNLI & Q$\to$G & 0.839 & 126 & CNN/DM & Q$\to$G & 0.449$^*$ \\
63 & MNLI & Q$\to$D & 0.692 & 127 & CNN/DM & Q$\to$D & 0.474$^*$ \\
64 & MNLI & Q$\to$Q & 0.836 & 128 & CNN/DM & Q$\to$Q & 0.988 \\
\bottomrule
\end{tabular}
\caption{All 128 scenarios. C=Claude, G=GPT, D=DeepSeek, Q=Qwen. $^*$Failure (AUC $\leq 0.50$): 11 total, in CUAD (5), SST-2 (3), CNN/DM (3).}
\label{tab:all_scenarios}
\end{table*}

\end{document}